# The non-contact assessment of the bridge expansion joint

**Pavel Ryjáček**

*Faculty of Civil Engineering CTU in Prague, Prague, Czech Republic*

**Kirill Golubiatnikov**

*Faculty of Civil Engineering CTU in Prague, Prague, Czech Republic*

**Ondřej Simon, Libor Ládyš, Martin Ládyš, Ondřej Červenka**

*EKOLA group, Prague, Czech Republic*

**Petr Klokočník**

*Statotest s.r.o., Liberec, Czech Republic*

**Contact:** pavel.ryjacek@fsv.cvut.cz

## Abstract

The aim of the paper is to provide information on a newly developed design methodology for the evaluation of bridge expansion joints with respect to their noise emission and overall technical condition. The methodology also gives recommendations for operational non-destructive measurements of the condition of bridge gates using a crossing laser sensors, CPX sensors or the necessary technical equipment and the method of collecting, processing and evaluating the measured data. The method can easily scan and evaluate geometry and/or noise emission of extensive number of joints by passing them without traffic interruption. The aim is to establish a methodology for comparing the noise and condition of bridge expansion joints in the road network, both over time (monitoring long-term trends in noise emission and degradation) and comparing different states and different types of bridge closures between each other.

## 1 Introduction

In the scope of acoustics, bridge expansion joints (EJ) are one of the one of the most problematic parts of the infrastructure. Vehicle crossings over the EJ cause a specific acoustic manifestation of impact or even impulsive character, which, in addition to increase in the noise load in the surroundings, has a relatively significant subjective annoying character for the nearest buildings. The intensity of these manifestations depends both on the type, quality and gap of the EJ itself and on the transition area of the EJ connection to the road surface. The quality of this transition proves to be a significant factor in the overall noise emission of the EJ.

The overall acoustic performance of the EJ consists of two main components:

1. the upper (surface) noise component from the tyre passing directly from the surface of the EJ itself, but also from the surface of the leading and trailing edge of the road.
2. The lower noise component - when a tyre passes over the surface of the EJ itself, an acoustic wave is generated in the semi-closed space of the expansion joint and is radiated downwards, or sideways if the gap is opened to the side.

One or the other component always becomes dominant in the vicinity of the EJ depending on the location of the control point. As the distance increases, the individual components lose their dominance, they can no longer be distinguished and only their cumulative influence begins to act.

The paper describes the developed solution of the non-direct measurement of the geometry of the EJ and the gap, and also the noise measurement, that can be done in the mass processing.

## 2 Expansion joints in Czechia

During the project, the detailed and extensive database was created, based on the EJ in Czechia. From the database, different types of graphs were then compiled for statistical processing. The database includes:

- total number of bridges - 2 296
- total number of EJ - 4 540
- from that, EJ with noise reduction - 247
- total length of EJ - 64 920 m
- average expansion length - 110 m

The most commonly used type of EJ the single gap joint (with a total length of 35 700 m). The age of this EJ is most often around 10-15 years and newer, nevertheless, there are still EJs that are 50-60 years old. According to Figure 7 we can observe an increasing trend in the number of noise-reduced EJ on the highways.

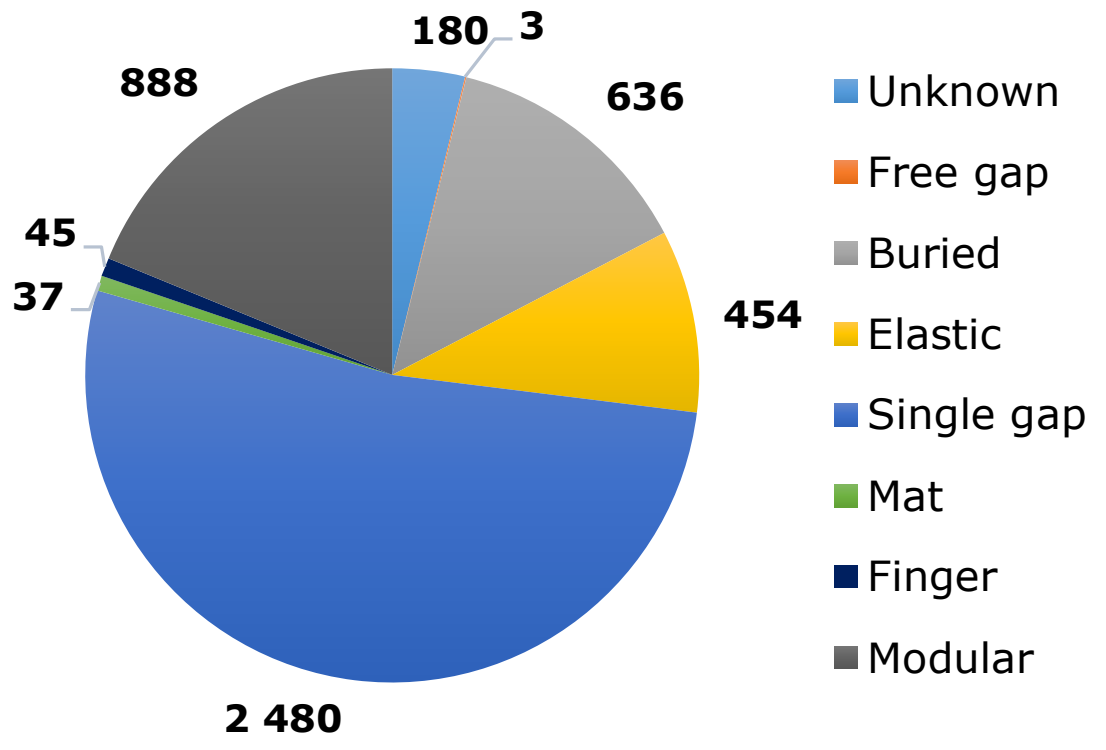

*Figure 1 Distribution of the EJ on the highway network*

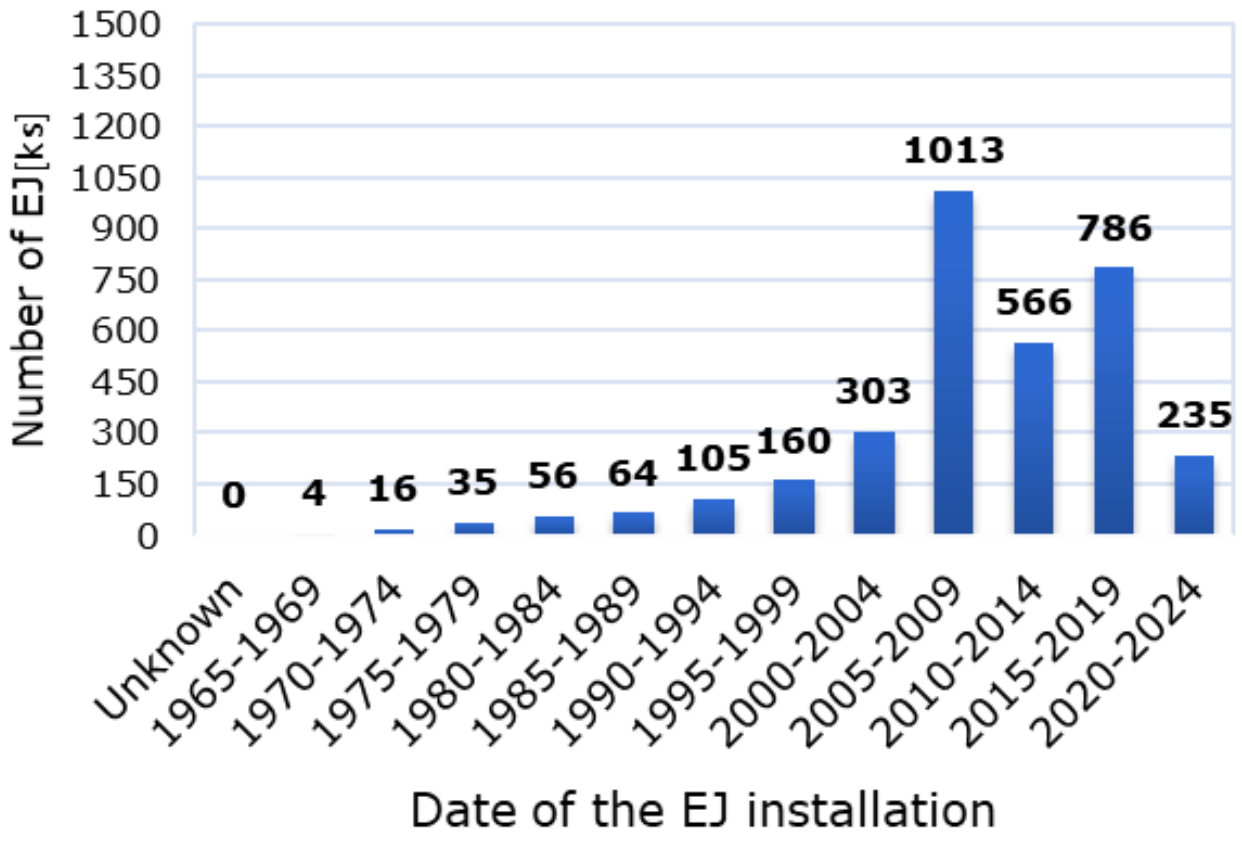

*Figure 2 Date of the EJ installation on the highway network*

As part of the assessment, there are also graphs showing the current state of the bridge gates in terms of their reliability. According to the Figure 3, only 9.2% of the EJ are in poor condition, while more than 64% are without visible defects. The condition of the transition area is more favourable. Almost 75% of the transition areas are completely free of defects and only 24% of the transition zones contain defects of various types (pothole, unevenness). Figure 5 shows the condition of the EJ with regard to the year of installation. The graph shows the expected trend that the older the joint, the worse is the condition.

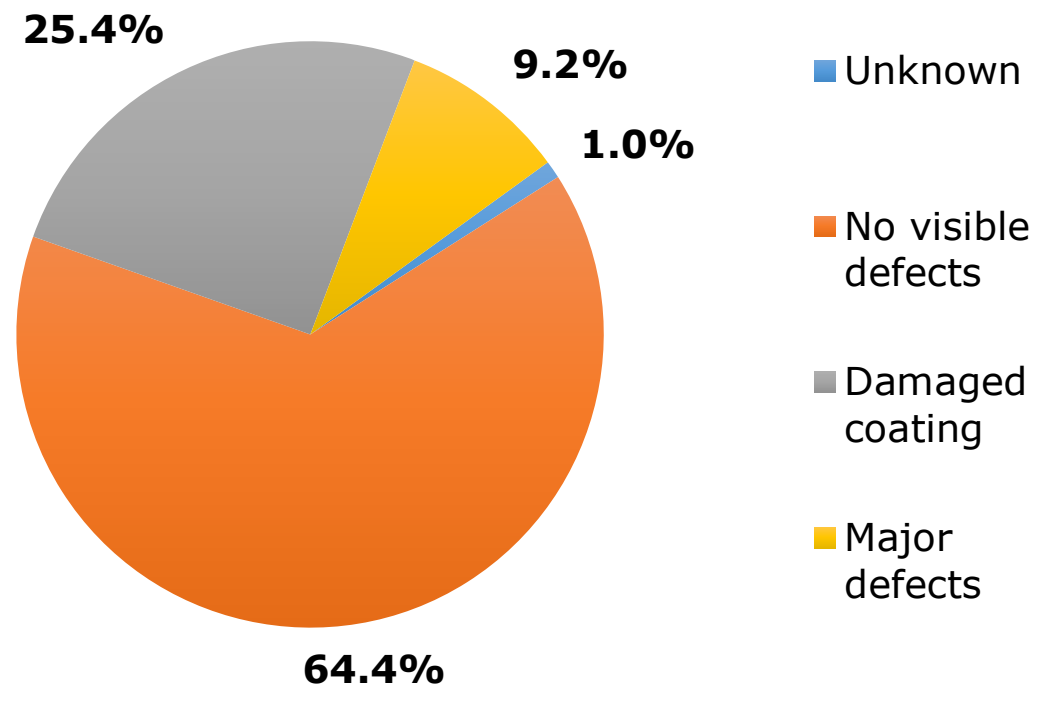

*Figure 3 Damage level of the EJ on the highway network*

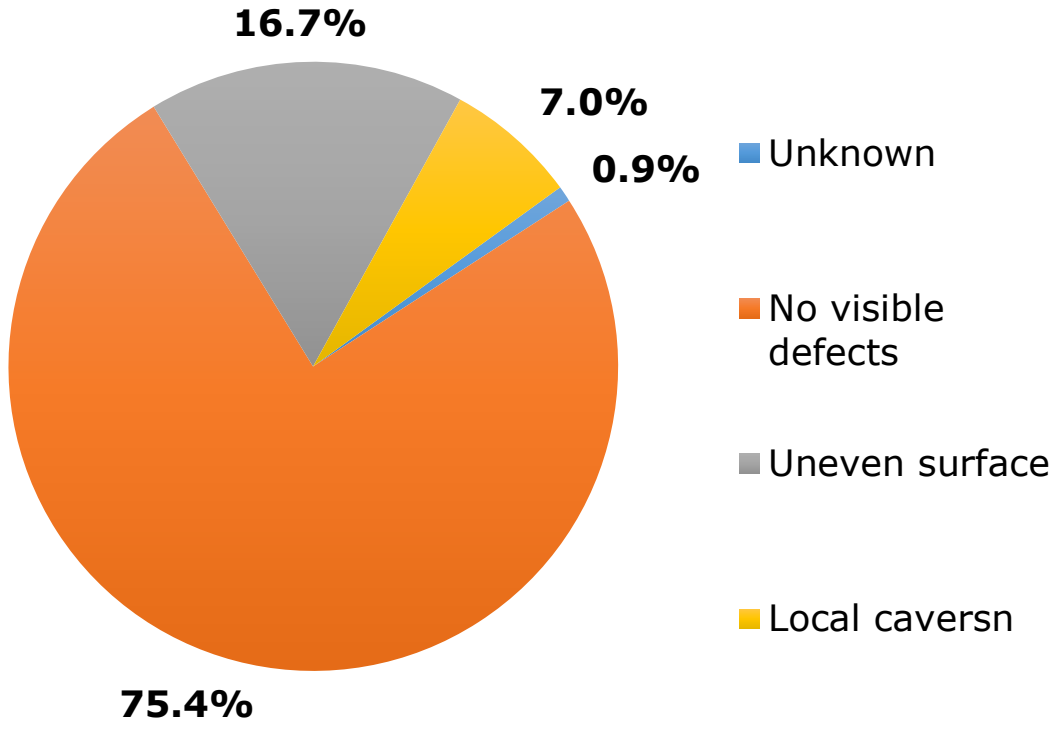

*Figure 4 Damage level of the transition zone on the highway network*

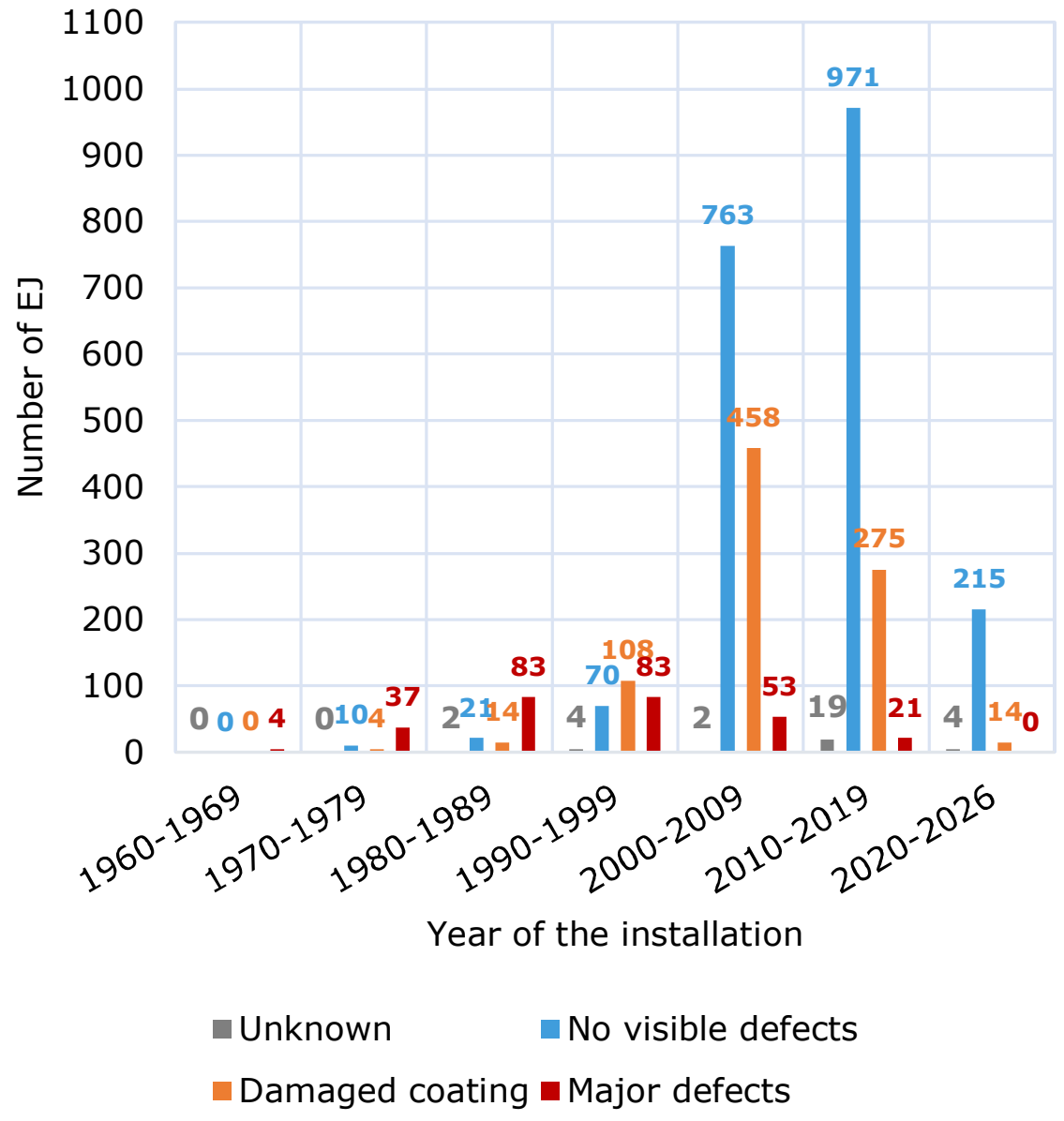

*Figure 5 Damage level of the transition zone on the highway network*

# 3 Measuring the expansion joint geometry

## 3.1 Traditional measurement

The good geometry of the EJ is essential for the fluent traffic and good level of the noise. To check it, and also to check the temperature movements of bridges, the measurements examine deviations in flatness, EJ inclinations (lateral and longitudinal) and EJ gap at a certain measured temperature. The EJ opening is measured using a displacement gauge. The temperature of the air, the superstructure and the girder itself is also measured. Inclination and deflections are first examined using a 4 m calibrated bar, which is placed longitudinally in the traffic lane.

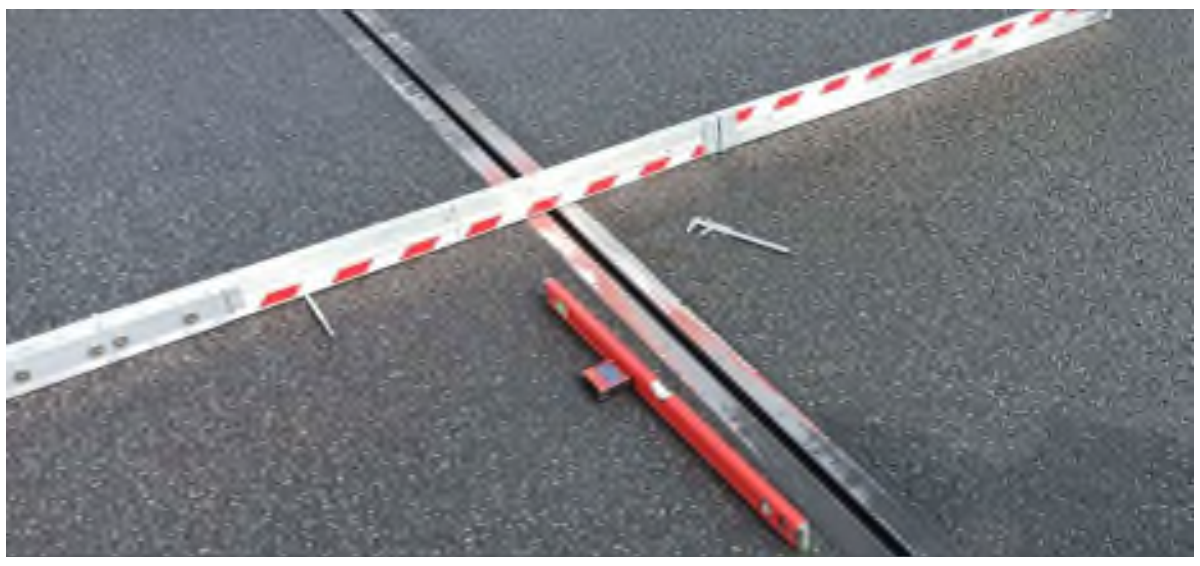

*Figure 6 The view on the EJ measurement*

This measurement is traditional and the results are generally good, the main problem is, that the measurement requires closure of the traffic lane, placement of the traffic signs that can create congestions. Also, interruption of the traffic is always a risk for the inspectors. Because of that, the new methods that can avoid that are needed.

## 3.2 Proposed laser scanning method

In addition to the measurement of the flatness of the bridge girder described above, the optical method was developed and tested. For this purpose, a specialised optical system was used to measure surface roughness and flatness. The system works on the principle of a suitable integration of an optical distance sensor - laser and a precision radar, the combination of which is able to measure and subsequently reconstruct the structure of the running surface at high speeds and resolutions - up to 1 mm resolution at 100 kph.

This system is part of a special measurement trailer for measuring road surface noise using the Close Proximity Method (CPX), which is owned by authors.

In the assessment of the EJ, the data obtained with this system are considered as the main source of information on the geometric shape of the EJ, since the measurements by means of the passage of the measuring trailer do not require significant traffic restrictions due to their speed, and the data collection with this system can therefore be carried out flexibly and efficiently even in the case when a complete closure of the lane is not feasible. Fully automated geometry measurement using the optical system is then one of the prerequisites for the successful completion of the planned large-scale technical screening on expansion joints.

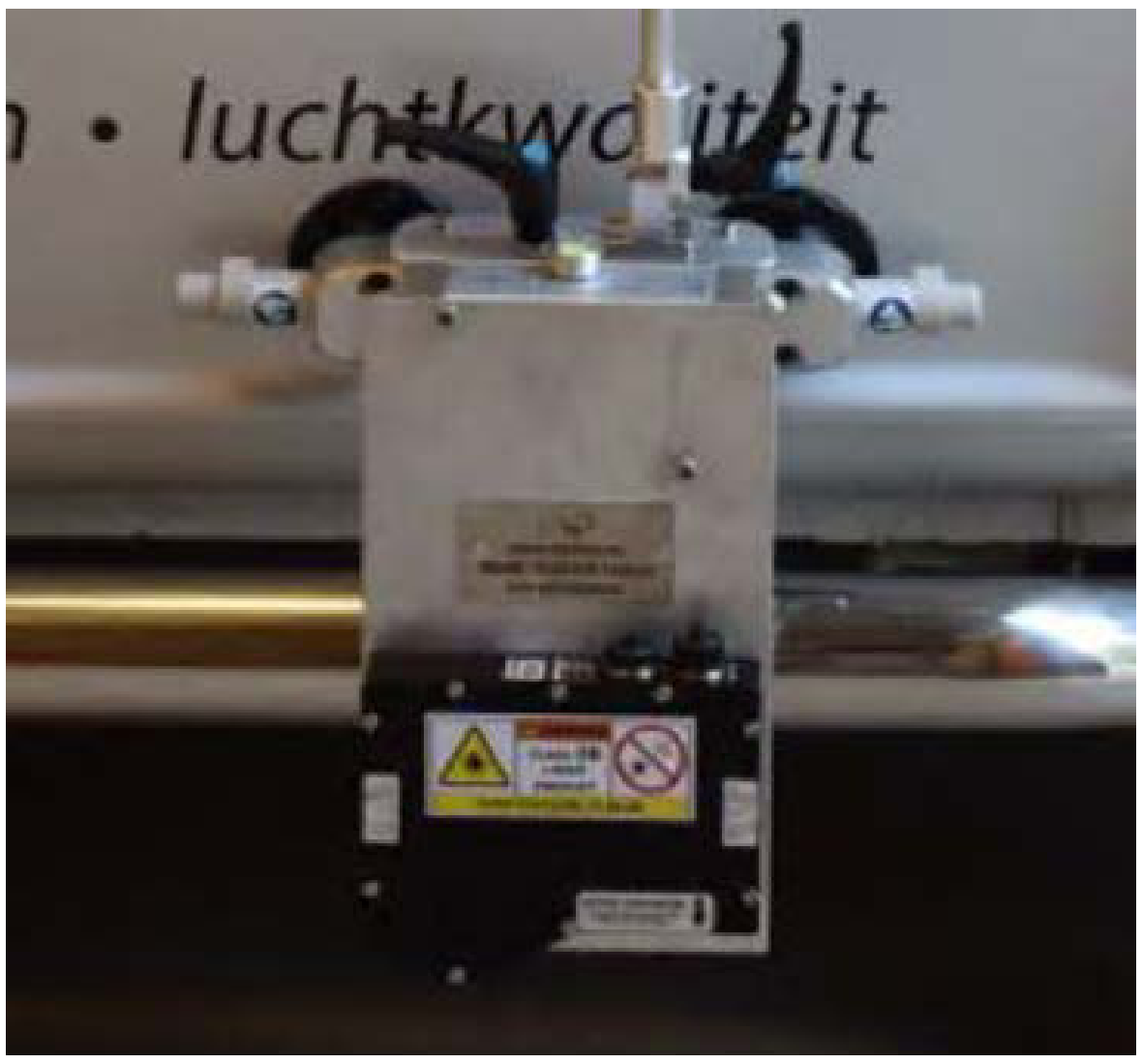

*Figure 7 The view on the optical system*

From the practical point of view of the applicability of optical measurements during the movement of the measuring trailer, it is a suitably chosen compromise between the travelling speed (with regard to the actual speed of the traffic flow and the influence of speed on the emission noise of the EJ) and the technical limits of the optical system - the sampling frequency and thus a sufficiently good accuracy of the reproduction of the geometric profile with respect to the speed. In order to establish these boundary conditions for valid optical data acquisition, measurements were made during repeated passes at different speeds. A sample of the measured data is shown in the following figure.

However, for the purposes of the system design of the criterion method of EJ evaluation, the zoomed part of the EJ itself and the immediate surroundings is of particular interest. The profile obtained there as a function of the passing speed is shown on the Figure 9.

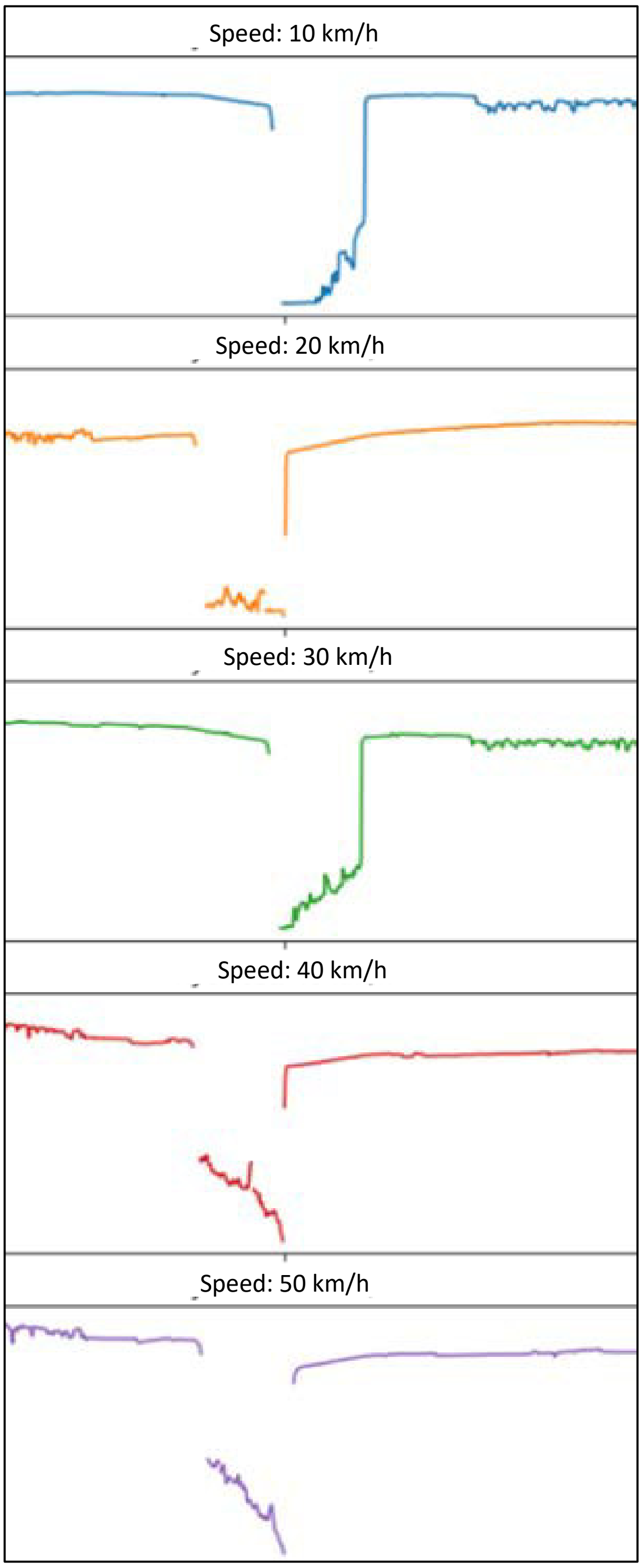

*Figure 8 EJ Transition with speed 10-50 kph*

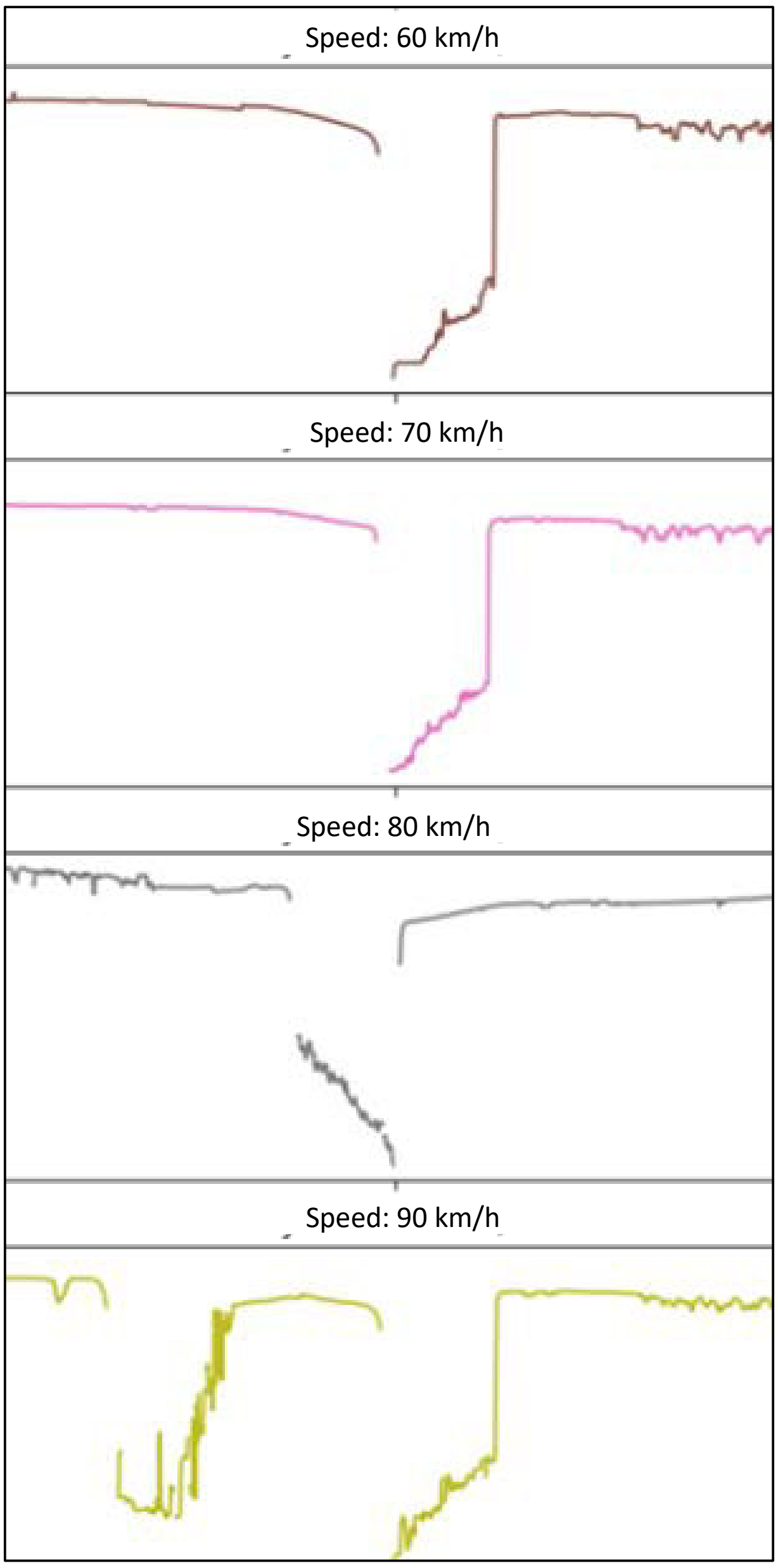

*Figure 9 EJ Transition with speed 60-90 km/h*

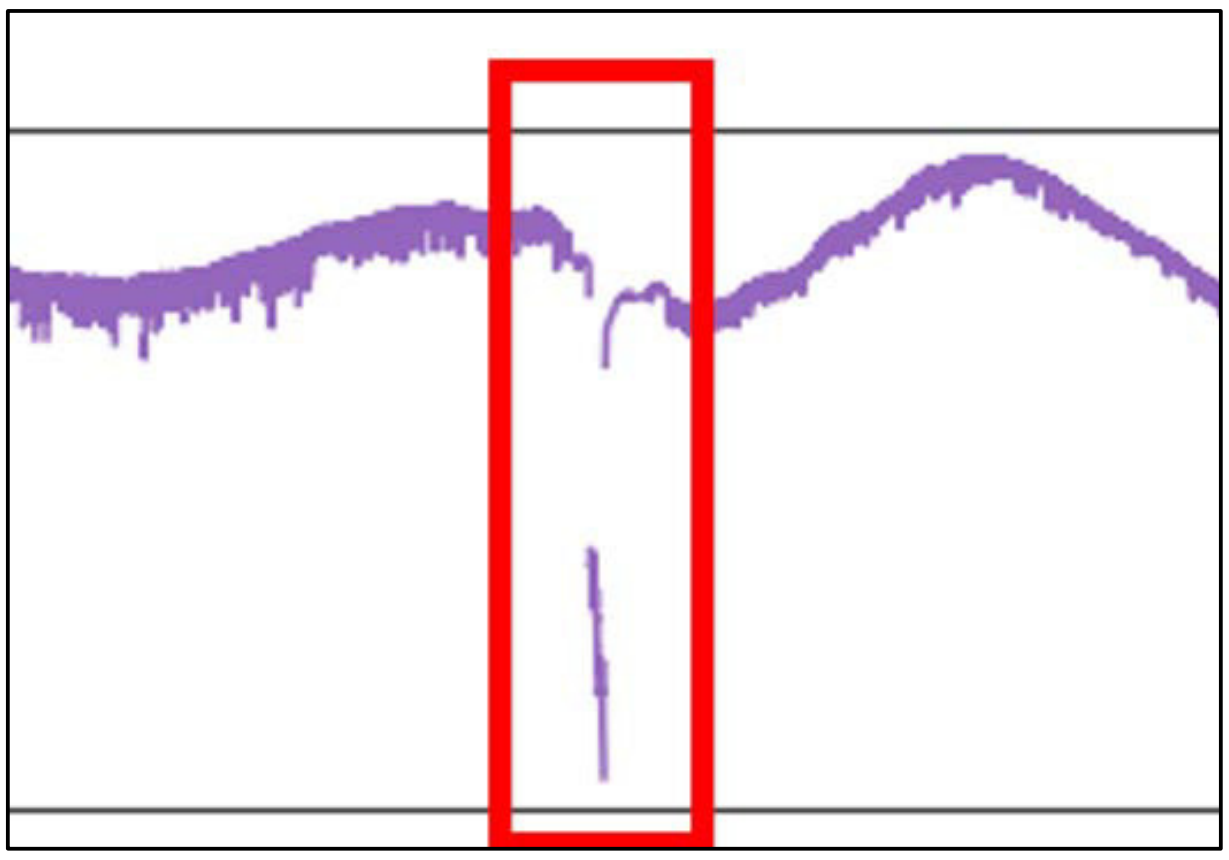

*Figure 10: Detail of the transition - 50 kph*

For a more detailed analysis, let's focus again on one particular passage and for a better analysis of the individual parts of the obtained profile, let's project the individual parts into the specific situation of the EJ.

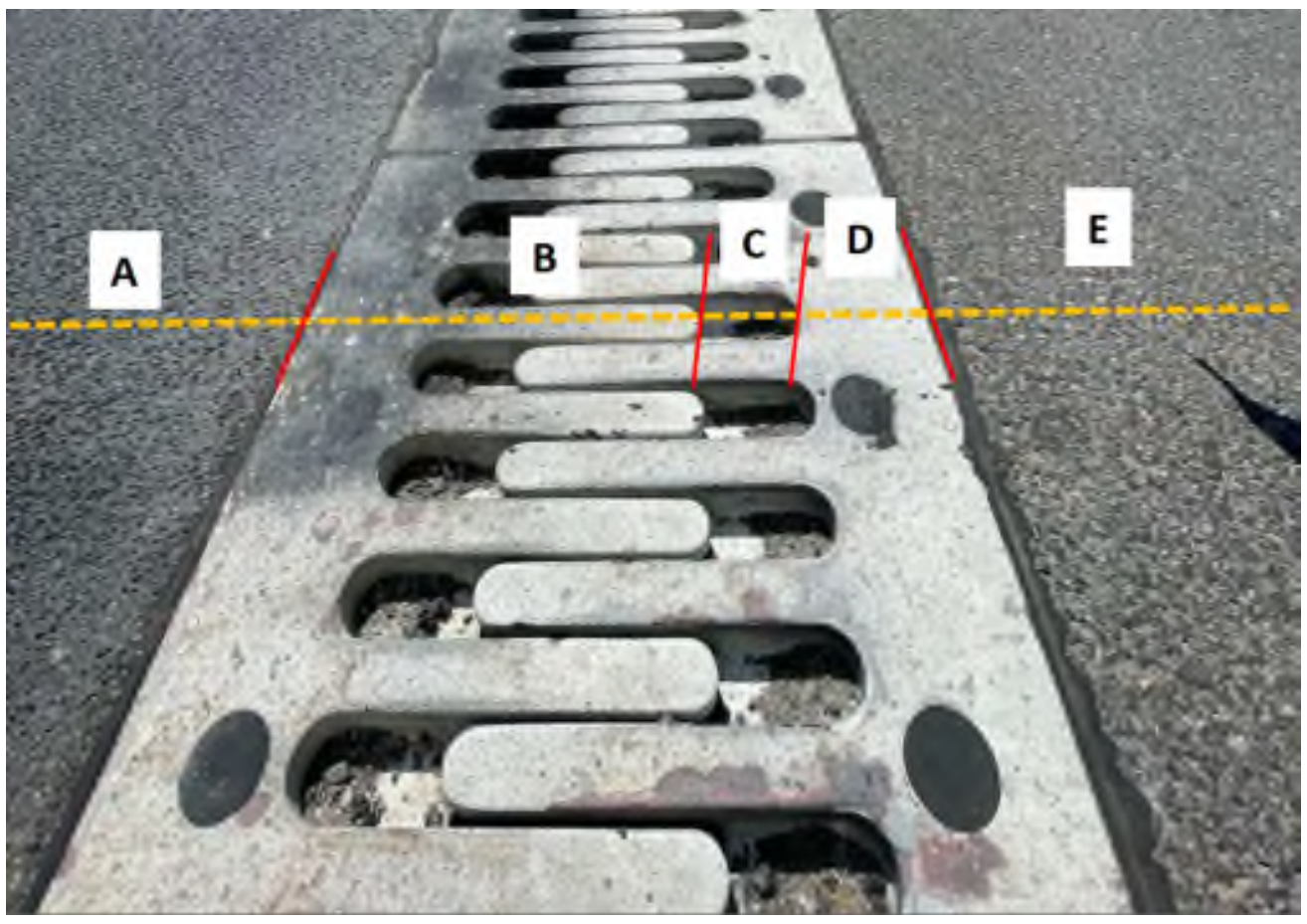

*Figure 11 Description of the EJ parts – finger joint*

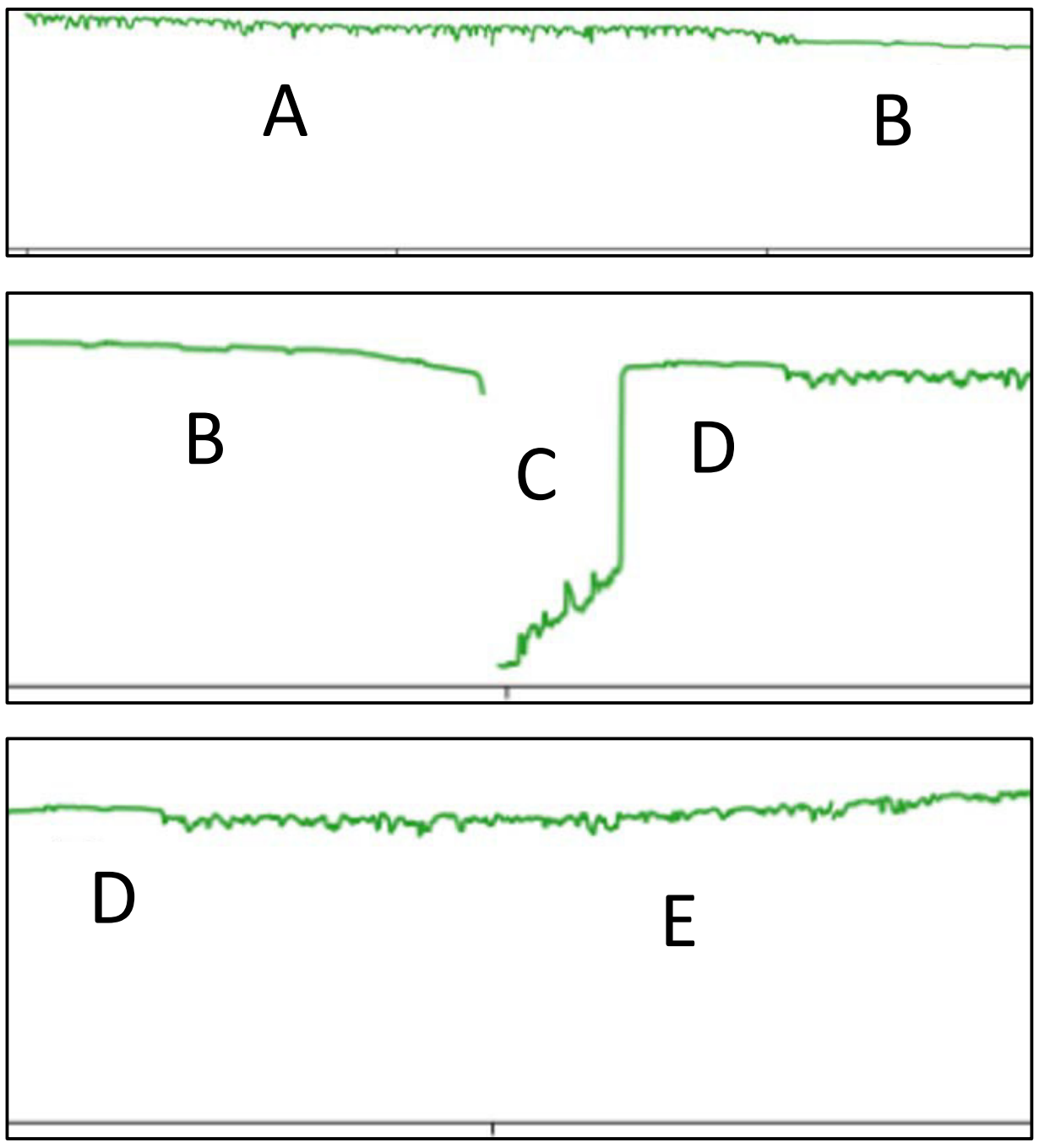

*Figure 12: Detail of the transition - 50 kph*

A - transition area before the EJ

- Here you can see the well reconstructed structure of the pavement surface with a gradual slope towards the EJ

B - the first part of the MZ with the "finger"

- At the beginning of B, a small stepped deepening of the EJ is visible compared to the level of the of the transition section, almost flat profile confirms the real smoothness of the surface part of the MZ
- The design of the EJ respects and continues the natural slope of the road

C - expansion joint - gap

- In this part, the laser detects the EJ, which is placed at 4 cm below the surface, the dirt deposits in the joint at the second part of the EJ are also clearly visible here

D - the second part of the EJ

- Second part of the EJ, shorter due to the tooth against the "finger"

E - transition area behind the end

- The total drop of the transition area by about 20 mm is visible, which confirms the finding of the geometric measurements

For the 90 kph passage, the measuring beam most likely "hit" exactly between the two fingers of and created a "fictitious gap" in the middle of the EJ during the transition from one finger to the other. Even here, however, the reproduction of the profile is very good.

On the basis of the above evaluated, it can be stated that the system in its current configuration is capable of a very good reproduction of the geometry of the bridge girder even at the highest tested travelling speed of 90 km/h and in terms of efficiency and sufficient accuracy is suitable for use as the main source of geometric data for the planned large-scale technical screening of EJ.

## 3.3 CPX noise measurement

Together with the laser geometric profile measurement, the $L_{CPX}$ parameter (dB) was recorded during the individual passes of the measuring trailer, which represents the sound pressure level of the road emission noise measured according to the EN ISO 11819-2 "Acoustics - Measurement of the effect of road surfaces on traffic noise - Part 2: Short distance method".

The CPX method is measured using a set of microphones placed in close proximity to the wheel with a special reference tyre. Thanks to the appropriate design of the trailer, this measurement configuration allows the noise of the rolling itself and other acoustic phenomena occurring through the interaction of the measuring tyre with the running surface to be recorded, without any disturbance from the surroundings (aerodynamic noise, surrounding vehicles, etc.).

The following figure shows the measurement set-up of a vehicle with a trailer at the location of the Hvězdonice bridge.

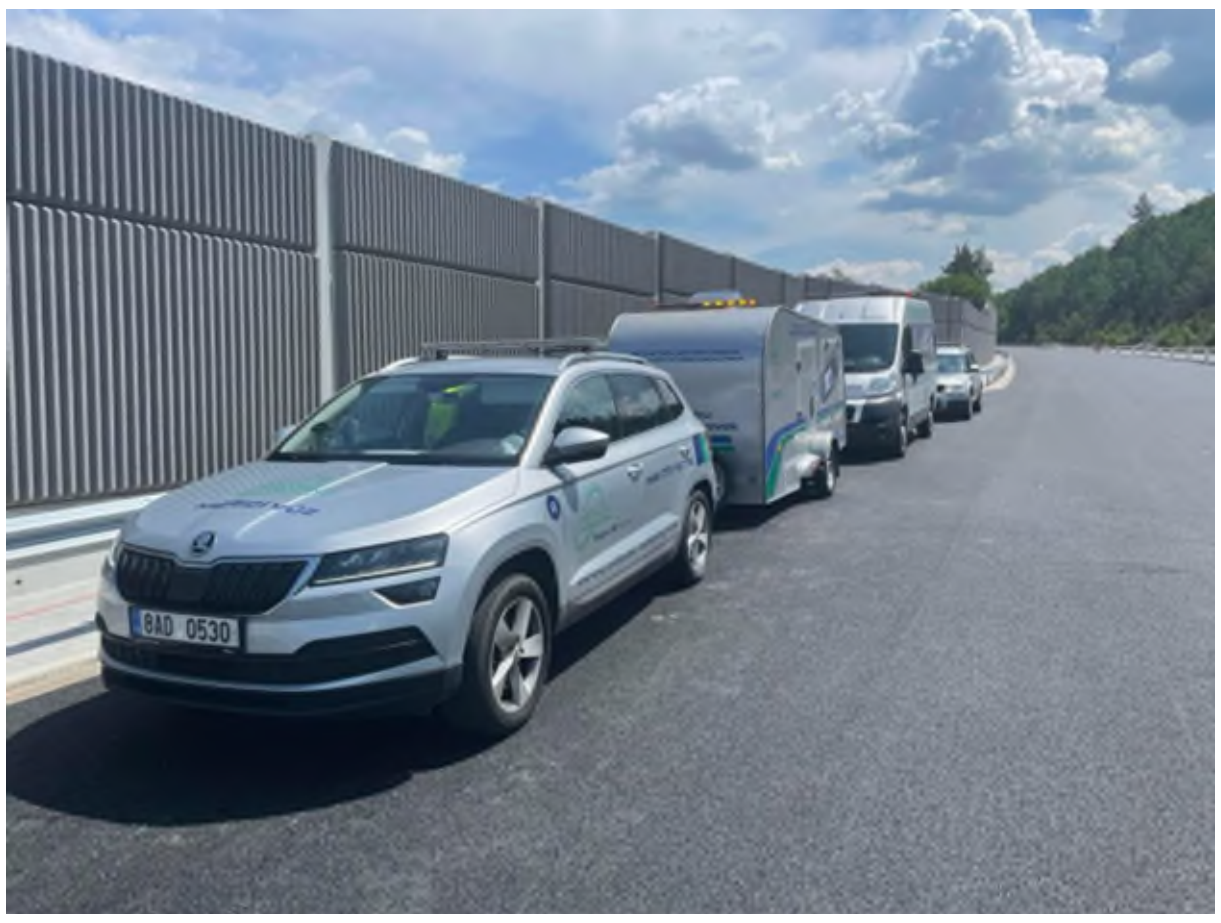

*Figure 13 Measurement trailer CPX*

The measurements were made for speeds of 10-100 kph and as part of the evaluation, the bridge closure and transition area section was evaluated separately from the road surface in front of the closure. Detailed results are compiled in the relevant research report, and for the purposes of this document the results for the 30, 50 and 80 kph crossing speeds are presented below:

*Table 1. CPX method results*

| | $L_{CPX}$ (dB) | | |
|---|---|---|---|
| **Speed [kph]** | **Road** | **EJ** | **Difference** |
| 30 | 82,0 | 83,2 | +1,2 |
| 50 | 91,2 | 91,6 | +0,4 |
| 80 | 97,7 | 98,4 | +0,7 |

## 3.4 Expansion joint passport

For the purpose of further technical screening, the format of the "EJ Passport" was developed, which presents a comprehensive overview of the measurement methods used and the most important results. The following figures show an example of the format of the processed passport.

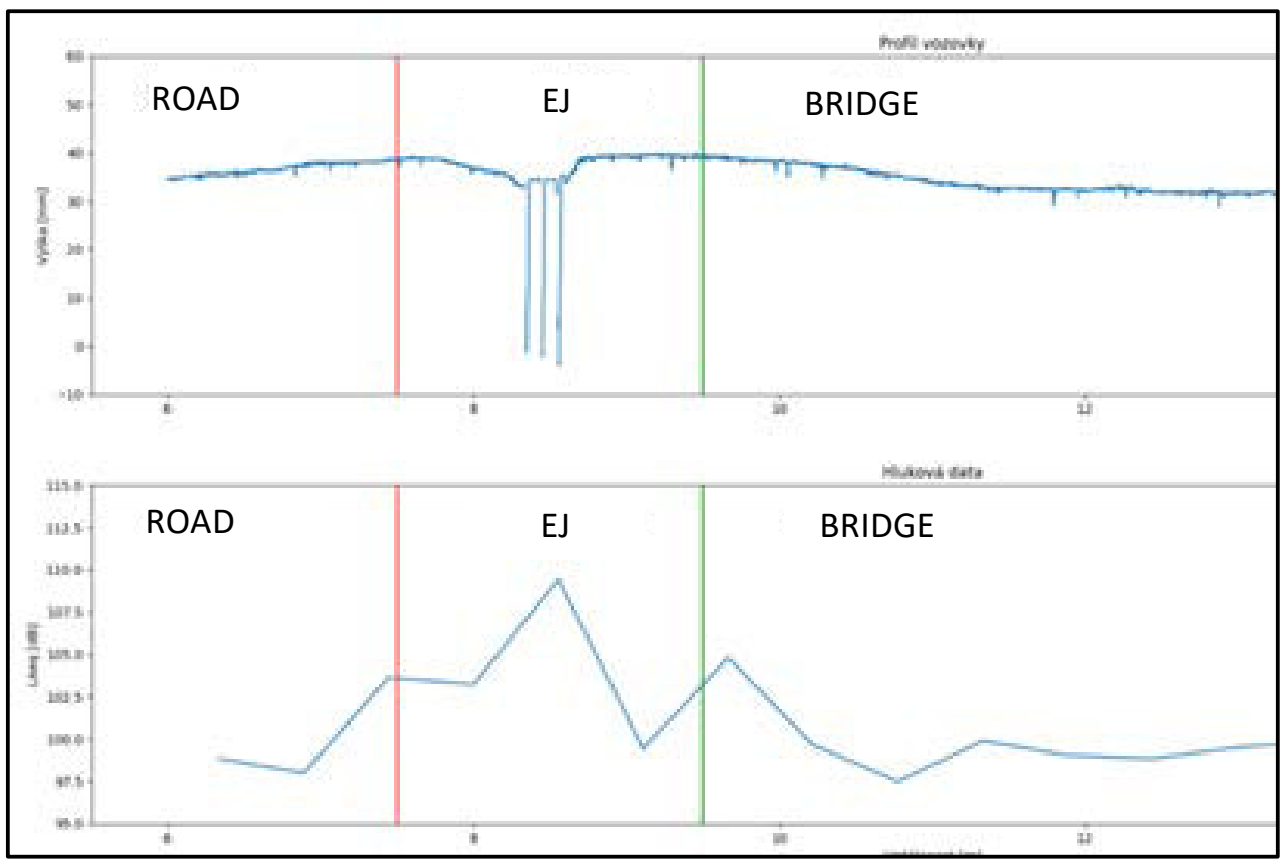

*Figure 14 EJ shape in the EJ passport – modular EJ*

## 4 Online image detection of the expansion joint defect

In case of defects, found on the EJ, the methods for the monitoring of the defects are necessary in order to prevent sudden failure of the EJ. The monitoring is extremely challenging – the acceleration on the EJ is reaching up to 500 m/s$^2$, the space is limited to the gaps only few centimetres wide. As a proposed and developed solution, camera IoT system was tested on an actual crack in the EJ in Prague highway bridge. The monitoring is performed using two camera boxes with a total of six cameras, the challenge here is mainly the correct lighting, the small distance and the dynamic effects on the unit. After analysing the possible cameras, fixed shutter stereo cameras and endoscopic cameras appeared to be optimal. The system is currently being tested in real operation.

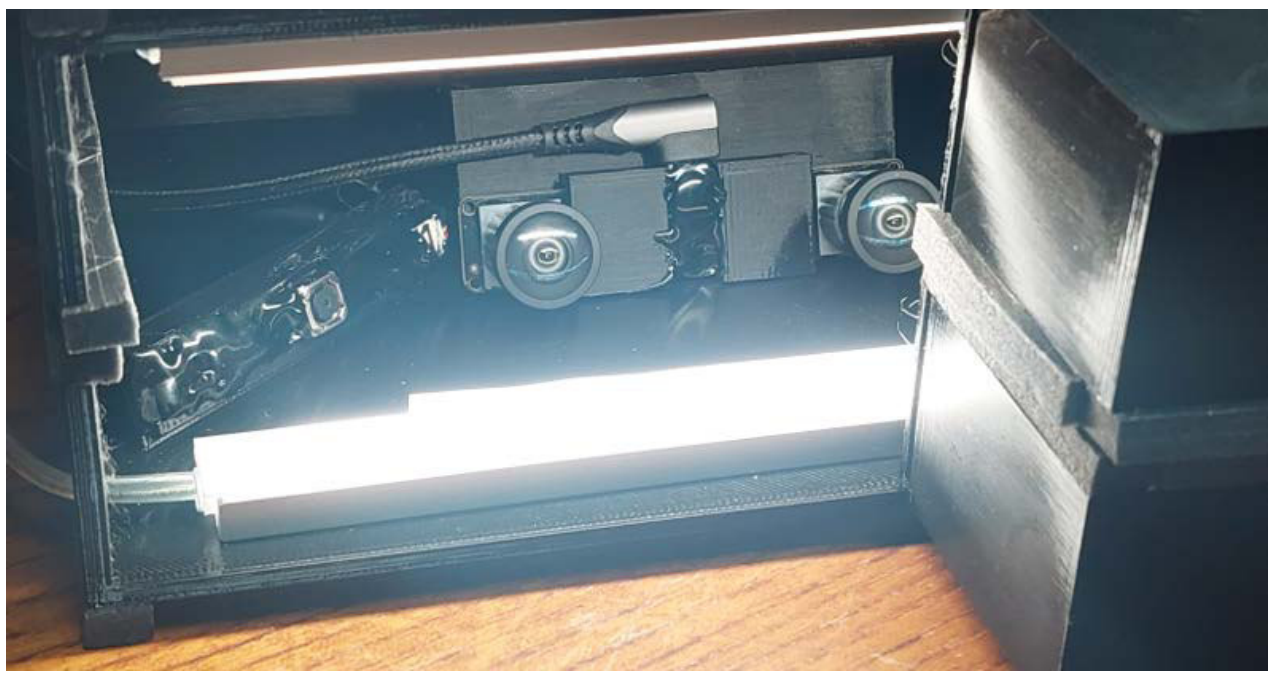

*Figure 15 Camera box before mounting on the EJ*

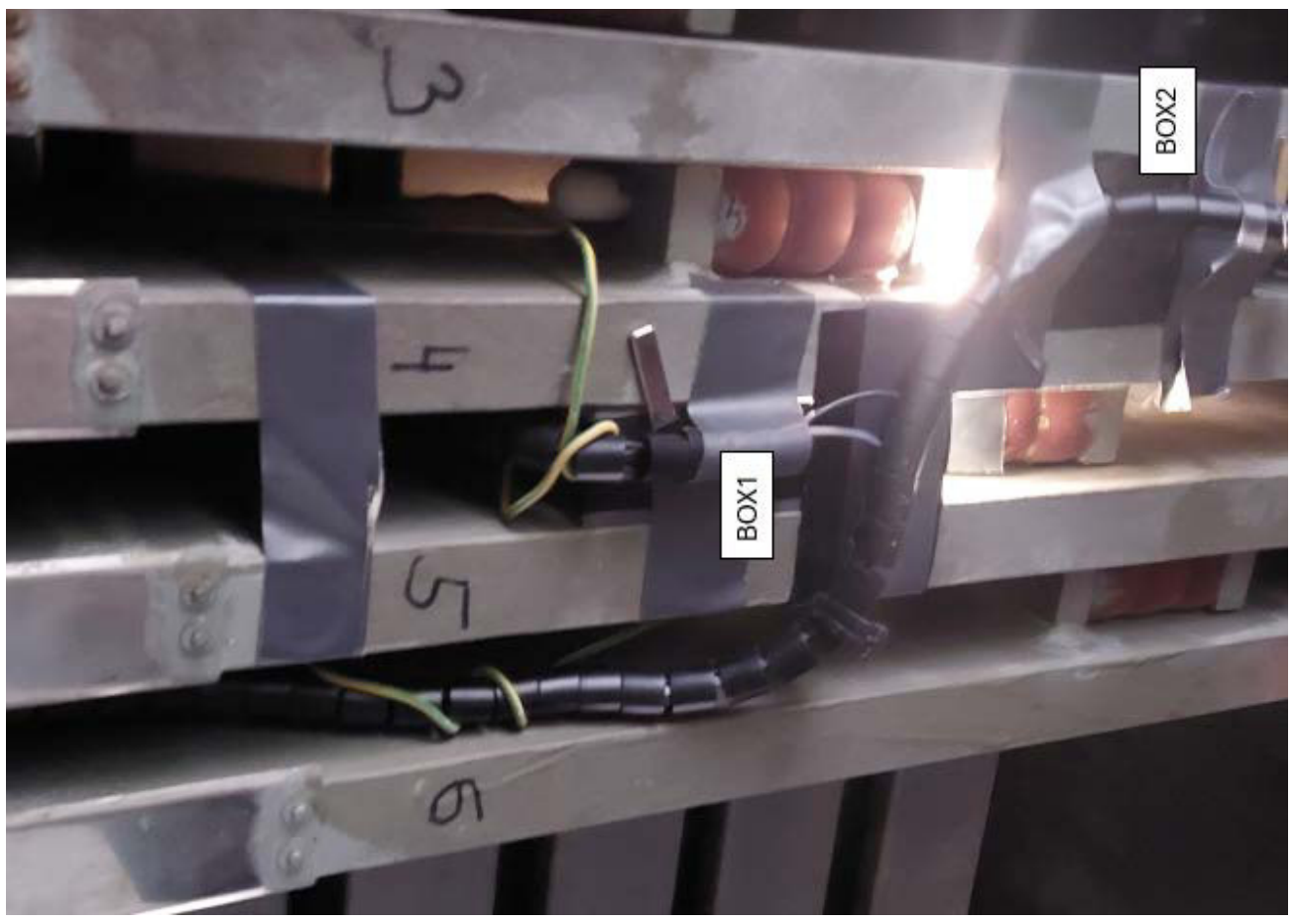

*Figure 16 EJ with camera boxes after installation*

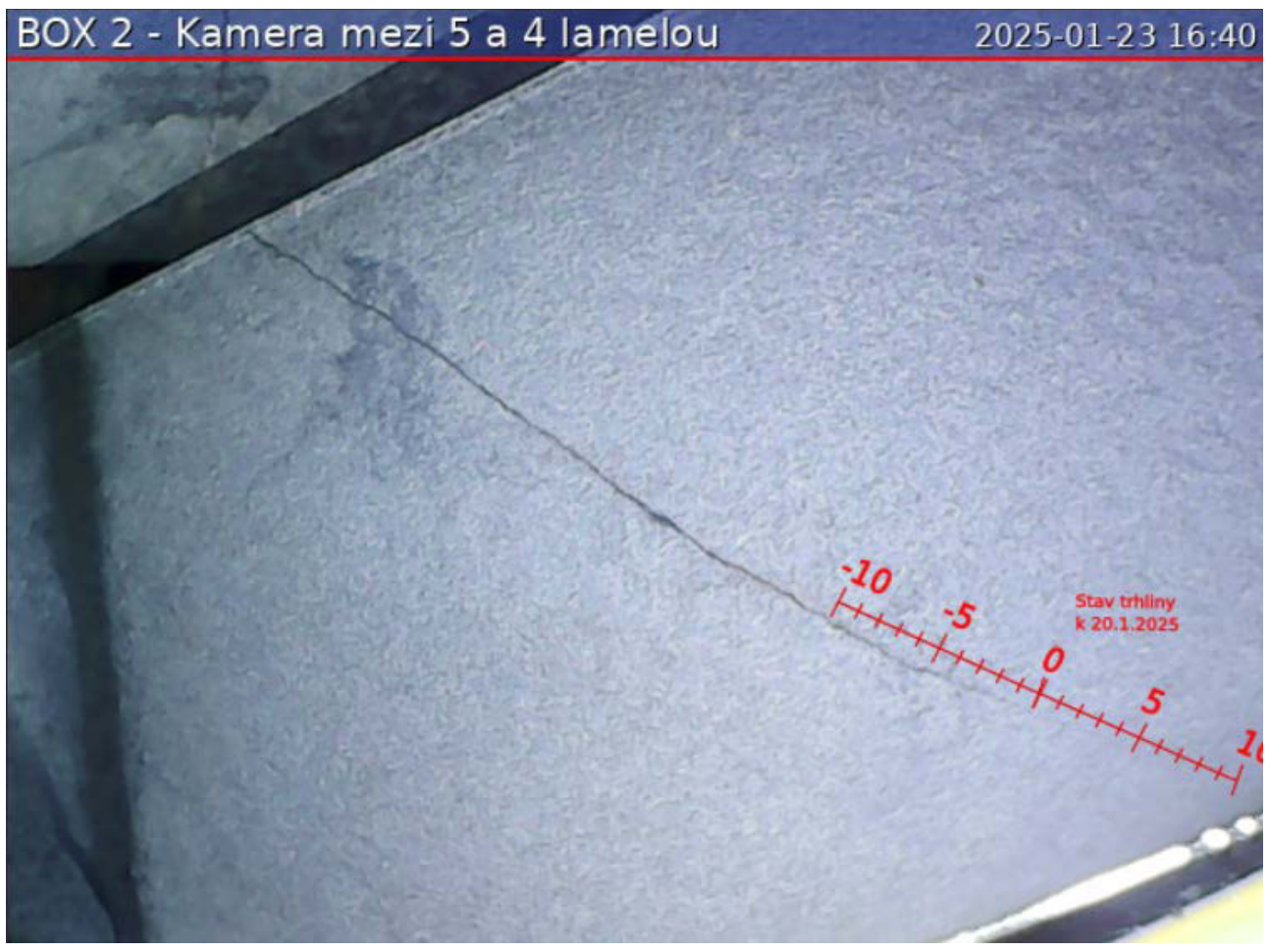

*Figure 17 Crack in the EJ beam – ruler shows the distance in mm and gives the time stamp*

## 5 Conclusions

The paper describes new possibilities to detect the expansion joint geometry, noise – and both in the fast way without traffic interruptions. Also, the camera IoT system were successfully developed and tested on a real bridge ensuring the safety before the crack reaches the critical length.

## 6 Acknowledgements

*The paper is partly funded by the project No. CK02000304 Criterial method for evaluating the noise emission of expansion joints after installation funded by Technological agency of Czech Republic.*

*The paper is partly funded in the chapter 4 by the project No. CK04000042 Identification and monitoring of progressive corrosion and non-*

*corrosion deterioration of steel bridges using IoT platform funded by Technological agency of Czech Republic.*

*Authors gratefully acknowledges financial support from project INODIN (CZ.02.01.01/00/23 020/ 0008487) co-funded by European Union.* 

## 7 References

[1] Lima, J., M., Brito, J.: Inspection survey of 150 expansion joints in road bridges. Engineering Structures 31 (2009) 1077_1084

[2] Foglar, M., Göringer, J.: Influence of the structural arrangement of bridges on the noise induced by traffic. Engineering Structures 56 (2013) 642–655